\documentclass[a4paper, amsfonts, amssymb, amsmath, reprint, showkeys, nofootinbib, twoside,aps,pre]{revtex4-2}
\usepackage[english]{babel}
\usepackage[utf8]{inputenc}
\usepackage{appendix}
\usepackage{graphicx} 
\usepackage{tikz}
\usetikzlibrary{arrows.meta,backgrounds,positioning,fit, shapes.geometric}
\usetikzlibrary{shapes.misc}
\usepackage{forloop}
\usepackage{xintexpr}
\usepackage{calc}

\usepackage[plain]{algorithm}
\usepackage{amsmath}
\usepackage{amssymb}
\usepackage{bm}
\usepackage[nolabel]{showlabels}
\usepackage[ISO=false]{diffcoeff}
\usepackage{listings}
\usepackage{graphicx}
\usepackage[utf8]{inputenc}
\usepackage{tablefootnote}
\usepackage{threeparttable}
\usepackage{makecell}
\usepackage{xcolor}
\definecolor{mydarkgray}{gray}{0.2}

\usepackage{color}

\usepackage{graphics}

\tikzset{cross/.style={cross out, draw=black, minimum size=2*(#1-\pgflinewidth), inner sep=0pt, outer sep=0pt},
cross/.default={1pt}}

\newcommand{\avg}[1]{\left\langle{#1}\right\rangle}
\newcommand{\qt}{{q_{\theta}}}

\renewcommand{\v}[1]{\mathbf{#1}}

\newcommand{\tr}{\operatorname{Tr}}

\usepackage{graphicx}
\usepackage{dcolumn}
\usepackage{bm}
\usepackage{textcomp}
\usepackage{color}
\usepackage{amsmath}
\usepackage[ISO=false]{diffcoeff}

\begin{document}

\title{Rényi entanglement entropy of spin chain with Generative Neural Networks}

\author{Piotr Białas}
\email{piotr.bialas@uj.edu.pl}
\affiliation{Institute of Applied Computer Science, Jagiellonian University, ul. \L ojasiewicza 11, 30-348 Krak\'ow, Poland}
\author{Piotr Korcyl}
\email{piotr.korcyl@uj.edu.pl}
\author{Tomasz Stebel}
\email{tomasz.stebel@uj.edu.pl}
\affiliation{Institute of Theoretical Physics, Jagiellonian University, ul. \L ojasiewicza 11, 30-348 Krak\'ow, Poland}
\author{Dawid Zapolski}
\email{dawid.zapolski@student.uj.edu.pl}
\affiliation{Faculty of Mathematics and Computer Science, Jagiellonian University, ul. \L ojasiewicza 6, 30-348 Krak\'ow, Poland }

\date{\today}

\begin{abstract}
We describe a method to estimate Rényi entanglement entropy of a spin system, which is based on the replica trick and generative neural networks with explicit probability estimation. It can be extended to any spin system or lattice field theory.  
We demonstrate our method on a one-dimensional quantum Ising spin chain. As the generative model, we use a hierarchy of autoregressive networks, allowing us to simulate up to 32 spins. We calculate the second Rényi entropy and its derivative and cross-check our results with the numerical evaluation of entropy and results available in the literature.

\end{abstract}


\maketitle

\section{Introduction}

Quantum entanglement is a phenomenon that underlies many existing and potential applications like quantum cryptography, quantum communication, and quantum computing \cite{Horodecki2009}. The degree of the entanglement between two parts ($A$ and $B$) of a quantum system can be quantified using the  von Neumann {\em entanglement entropy} 
\begin{equation}
    S(A) = -\tr \rho_A\log \rho_A,
\end{equation}
where $\rho_A$ is the reduced density matrix, i.e.
given the full density matrix 
\begin{equation}
    \rho_{ij} = \frac{\langle i|e^{-\beta H}|j\rangle}{\sum_i \langle i|e^{-\beta H}|i\rangle}
\end{equation}
of a bipartite system divided into part $A$ and $B$, the reduced density matrix $\rho_A$ is obtained by tracing out the $B$ part:
\begin{equation}
    \rho_A =\tr_B\rho.
\end{equation}
The estimation of the von Neumann entanglement entropy requires the full eigenspectrum of the matrix $\rho_A$. Because the size of the   
Hilbert space grows exponentially with the size of the system, such calculations are notoriously difficult.
Some simplification can be obtained by quantifying quantum entanglement by the quantum Rényi entropy of order $n$,
\begin{align}
S_n(A) = \frac{1}{1-n}\log \mathrm{Tr}\, \rho_A^{n}.
\end{align}
The von Neumann entanglement entropy can be recovered in the limit of $n \rightarrow 1$. The advantage of Rényi entropy comes from the fact that one can employ the replica trick together with the path integral formalism to rewrite the trace of the power of the reduced density matrix as the ratio of partition functions of systems multiplicated in the time direction with appropriate boundary conditions \cite{Calabrese_2009}. From this point on, several methods can be applied to
access such partition functions. For instance, one can implement a numerical Monte Carlo sampling procedure \cite{PhysRevLett.104.157201,Humeniuk:2012xg}. Taking as an example the second Rényi entropy, i.e. $S_2(A)$, one has to estimate the ratio of the partition functions of two systems, with a twice time extent differing in boundary conditions. The main difficulty hides behind the fact that the sampling algorithm has to easily transfer configurations with one topology of boundary conditions to the other and back. Failure may result in large autocorrelation time and subsequently large statistical uncertainty on the estimated ratio of partition functions. The problem increases very quickly with the system size and with increasing entanglement entropy. Several techniques have been devised to remedy this problem and ensure the correctness of the outcomes \cite{PhysRevB.90.125105,Zhao_2022}.  A separate class of Monte Carlo algorithms is exploiting the Jarzynski equality in non-equilibrium statistics to relate the change in free energy to the ratio of partition functions \cite{PhysRevLett.124.110602,Zhao_2022,Bulgarelli:2023ofi}. Among the other approaches aiming at estimating the rate of entanglement,
very popular are methods that use tensor networks to approximate the ground state of the system and then find the maximal overlap of that approximated ground state with the most general separable state \cite{Eisert_2013,Okunishi_2022}. 

\begin{figure}
	\centering
\includegraphics[width=0.45\textwidth]{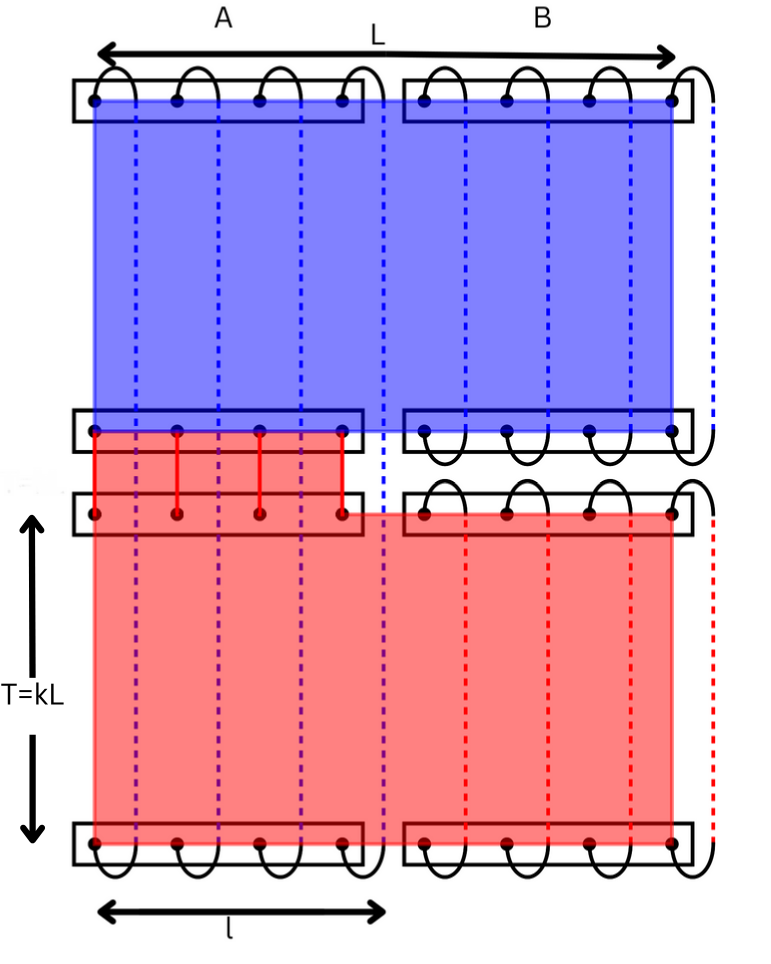}
	\caption{Sketch of spin division in replica trick. The two replicas are denoted by distinct colours: blue and red. $A$ and $B$ denote the two subsystems. $L$ is the physical length of the Ising chain, and $l$ the length of the subsystem $A$. $T$ is the length in the imaginary time direction of the single replica. The dashed lines show which spins have to be identified to form correct boundary conditions. }
	\label{replica_trick}
\end{figure}

In this contribution, we investigate an alternative approach. We do start with the replica trick and replicate the system in the time direction according to the prescription. However, we treat the numerical simulation of the original system and its replicated counterpart separately. We then access the partition functions of each of them independently by employing the recently proposed generative neural network architectures. Once correctly trained, the latter offer direct access to the partition function and other thermodynamic quantities of investigated statistical systems. Our aim is to provide quantitative estimates of the efficiency of such an approach and the possible precision of the results with the currently available neural network architectures. We use the quantum Ising spin chain in the transverse magnetic field as the test bench and study system sizes up to 32 spins, estimating the ground state bipartite second Rényi entanglement entropy as a function of $x=l/L$, where $l$ and $L$ are the lengths of part $A$ and the whole system $A\cup B$, respectively. 

The rest of the paper has the following structure. In Section \ref{sec. method} we briefly describe the details of the replica trick and provide the Hamiltonian of the Ising spin chain, introducing the required definitions of the partition functions. We also describe how generative neural networks based on autoregressive architectures can be used to sample configurations of the two-dimensional statistical spin system. In Section \ref{sec. results} we provide the overview of our results. In particular, we describe our results for $S_2(x)$ and its derivative with respect to $x$, $C_{n}(x)$, the so-called entropic $c$-function. We compare our results with those obtained through the Jarzynski inequality and off-equilibrium numerical simulations of Ref.\cite{Bulgarelli:2023ofi}. We conclude and provide some outlook in Section \ref{sec. summary}.

\section{Method}
\label{sec. method}

\subsection{Rényi entropy and replica trick}

Within the path integral formalism, the replica trick \cite{Calabrese_2009} allows to express the trace of an arbitrary power of the reduced density matrix $\rho_A $ in terms of a partition function of the system multiplicated in the imaginary time direction.
 More specifically, the quantization of a one-dimensional lattice system with $L$ spins gives rise to a discretized two-dimensional statistical system with dimensions $L \times T$. The division into parts $A$ and $B$ subsists for all imaginary times. We recover the quantized one-dimensional system in the limit $T\rightarrow \infty$ and $\epsilon \rightarrow 0$, where $\epsilon$ is the spacing in the imaginary time direction. Keeping $T$ finite corresponds to studying the quantum system at a non-zero temperature. Hence, for sufficiently large $T$ we expect that the dynamics of the quantum system is dominated by its ground state. The extracted Rényi entanglement entropy thus corresponds to the ground state entanglement entropy. In the implementation, we shall always set, $T=k L$ where $k$ is integer $\ge 1$, and investigate how the results change as we increase $k$.

In Fig.~\ref{replica_trick} we show schematically the topology of the duplicated system for $n=2$. Blue and red colours distinguish the two replicas, which are "glued" together at one imaginary time slice only for the subsystem $A$. The dashed lines show which spins have to be identified to form correct boundary conditions.   

For a given subdivision into parts $A$ and $B$ we denote as $Z_{n}(A)$ the partition function of the full system of replicated $n$ times and the standard partition function (of one replica) by $Z$. Then, the $n$-th Rényi entropy is given by \cite{Calabrese_2009}
\begin{align}
S_n(A) = \frac{1}{1-n}\log\frac{Z_{n}(A)}{Z^n}.
\label{Renyi_Z}
\end{align}

\subsection{Ising model}

In this manuscript, we consider the quantum Ising model in D=1+1 dimensions with the Hamiltonian:
\begin{align}
\label{Ising_ham_quantum}
\hat{H}=- J\sum_{\langle i, j\rangle} \hat \sigma_i^z \hat \sigma_j^z -h \sum_{ i} \hat \sigma_i^x,
\end{align}
where $\hat{\sigma}_i$ are spin operators and their standard representation in the $\hat{\sigma}^z$ basis is given by Pauli matrices. The second term is an interaction with an external (transverse) field. Using the path integral formalism, it can be translated to the two-dimensional classical Ising model without an external field \cite{Kogut1979}. For simplicity, we choose $J$ and $h$ such that in the corresponding classical model the couplings between the spins are equal in the imaginary time and spatial directions (together denoted as $\beta$). Therefore, the energy of the classical Ising model is given by: 
\begin{align}
\label{Ising_energy}
E({\v s})=- \beta \sum_{\langle i, j\rangle}  s_i s_j,
\end{align}
where $s=\pm 1$ and the periodic boundary conditions are used. Accordingly, the partition function $Z$ is given as:
\begin{align}\label{eq:Z}
Z = \sum_{\v s} e^{- E(\v s)}.
\end{align}

In order to calculate the partition function of the replicated system $Z_{n}(A)$, one needs to consider configurations of spins $\v s_{(n)}$ of the multiplicated system with interactions dictated by the replica trick. We therefore define the modified energy $E_{(n)}$, for which standard periodic boundary conditions are modified by the cut, as indicated in Fig.~\ref{replica_trick}. The partition function $Z_{n}(A)$ is then given by:
\begin{align}\label{eq:Z-n}
Z_{n}(A) = \sum_{\v s_{(n)}} e^{- E_{(n)}\left(\v s_{(n)}\right)}.
\end{align}

In order to eliminate the less interesting constant part and concentrate on the universal coefficient of the $l$ dependent part, it is customary to consider the derivative of $S_{n}$ w.r.t. $l$, so-called entropic c-function, which is given by \cite{Bulgarelli:2023ofi}:
\begin{align}    
    C_{n}(l)=\left[\frac{L}{\pi}\sin{\left(\frac{\pi l}{L}\right)}\right]^{D-1} \frac{1}{|\partial A|}\frac{1}{1-n} \times \nonumber \\ 
   \times \lim_{\epsilon \to 0}\frac{1}{\epsilon}\log{\frac{Z_{n}(l)}{Z_{n}(l+\epsilon)}}.
\end{align}
The term $|\partial A|$ corresponds to the boundary of the segment $A$, which for the one-dimensional system consists of two end-points of the segment, and hence is equal to 2. In practical calculations, we use the approximation of the derivative employing the second-order central finite difference formula:
\begin{equation}
    C_{n}(l)\approx \frac{L}{2\pi}\sin{\left(\frac{\pi l}{L}\right)}\frac{1}{1-n}\log{\frac{Z_{n}(l-\frac{1}{2})}{Z_{n}(l+\frac{1}{2})}},
    \label{c-funtion_approx}
\end{equation}
where we also have set $D=2$. In what follows, we shall calculate $ C_{n}(l)$ for fractional values of $l=3/2, 5/2, 7/2...$. Although, in this formula the standard partition function $Z$ has cancelled out, $C_{n}(l)$ still requires the estimation of the ratio of two distinct partition functions. 

\subsection{Neural Importance Sampling for partition functions}

In this contribution, we propose to estimate Rényi entanglement entropy by directly computing the partition functions \eqref{eq:Z} and \eqref{eq:Z-n} using generative neural networks
with explicit probability estimation.
We denote the probability distribution, modelled by the neural network, by $q_\theta$, where $\theta$ collects all the parameters of the neural network. It is important to distinguish it from the Boltzmann probability distribution, which is the target for training:
\begin{equation}
p(\v s) = \frac{1}{Z} e^{- E(\v s)}.
\label{eq. p}
\end{equation}

In what follows, we shall focus on autoregressive neural networks (ANNs) \cite{2019PhRvL.122h0602W}, because they are well adapted to discrete spin systems. ANNs have proven to be effective samplers of Ising model \cite{Bialas:2021bei,Bialas:2022qbs} and can be in principle applied to any model with discrete degrees of freedom \cite{Bialas:2022bdl}. However, it should be noted that our method is not restricted to any particular architecture. 

ANNs utilize the probability product rule, where $\qt$ is factorized into a product of conditional probabilities
\begin{equation}
\label{eq:factorisation}
    q_{\theta}(\v s)  =  \prod_{i=1}^{N_{spin}} q_{\theta}(s^i|s^1,s^2,\dots,s^{i-1}).
\end{equation}
One samples spin configurations from $q_\theta$ by ancestral sampling, i.e. by fixing the spins one after another based on the conditional probabilities and the values of previously fixed spins. 

The aim of the training is to tune the parameters $\theta$ so that $\qt$ is close to $p$ with respect to the Kullback-Leibler divergence:
\begin{align}
    D_\textrm{KL} (q_\theta | p) &= \sum_{\v s} q_\theta(\v s) \, \log \left(\frac{q_\theta(\v s)}{p(\v s)}\right).
    \label{eq:KL_loss} 
\end{align}
Once the network is trained, one can use it to sample configurations and calculate partition functions. We first note that Eq.\eqref{eq:Z} can be written as:
\begin{equation}
    Z=\sum_{\v s}\qt(\v s) \frac{e^{-\beta E(\v s)}}{\qt(\v s)}\equiv\avg{\hat w(\v s)}_{q_{\theta}},
    \label{def_Z_N}
\end{equation}
where we defined the weights $\hat w(\v s)={e^{-\beta E(\v s)}}/{\qt(\v s)}$ and the average of the r.h.s. is performed over the probability distribution $\qt$. The latter can be approximated by the standard arithmetic mean from $N \gg 1$ configurations, distributed according to the distribution $\qt$:
\begin{equation}
    Z \approx  \frac{1}{N}\sum_{i=1}^N \hat{w}(\v s_i)\equiv \hat Z_N,   \qquad \v s_i \sim q_{\theta},
\end{equation}
where by $\v s_i \sim q_{\theta}$ we mean that configurations $\v s_i$ are drawn from the distribution $q_{\theta}$.

Note that although the above formula formally does not require $\qt$ to be close to $p$, if this is not the case, the uncertainty of $\hat Z_N$ will be large. The procedure of sampling configurations from the approximate probability distribution provided by the neural network together with the estimation of the partition function was proposed in Ref.~\cite{2020PhRvE.101b3304N} and named Neural Importance Sampling (NIS).

In this work,  NIS is applied to quantum entanglement entropy for the first time. Since we evaluate each partition function separately, we avoid the problem of large ratios of partition functions, which in principle can render typical Monte Carlo estimators ineffective due to exponentially large autocorrelation times. 

\subsection{Numerical setup}

In order to improve the scaling of the numerical cost of the training of neural networks, we use the hierarchical algorithm proposed in Ref.\cite{Bialas:2022qbs}. Compared to the original autoregressive networks approach \cite{2019PhRvL.122h0602W}, the single autoregressive neural network is replaced by a recursive hierarchy of much smaller neural networks. We have adapted the geometry of these networks to the particularities of the replica topology. The details are described in Appendix \ref{han}.

The estimation of entanglement entropy Eq.\eqref{Renyi_Z} and its derivative Eq.\eqref{c-funtion_approx} requires two different partition functions for each value of $k$, $l$ and $L$. For the calculations presented below, we have independently trained neural networks for different values of $k=1, \dots, 8$, system size $L=8,16, 32$, and each value of $l=1,\ldots,L-1$. We profited from transfer learning by training the neural networks at $l=L/2$ and using the trained networks as the initial state for the training at $l=1$, $l=2$, and proceeding in this way until $l=L-1$.

\section{Results}
\label{sec. results}

\begin{center}
\begin{figure*}
    \includegraphics[width=0.75\textwidth]{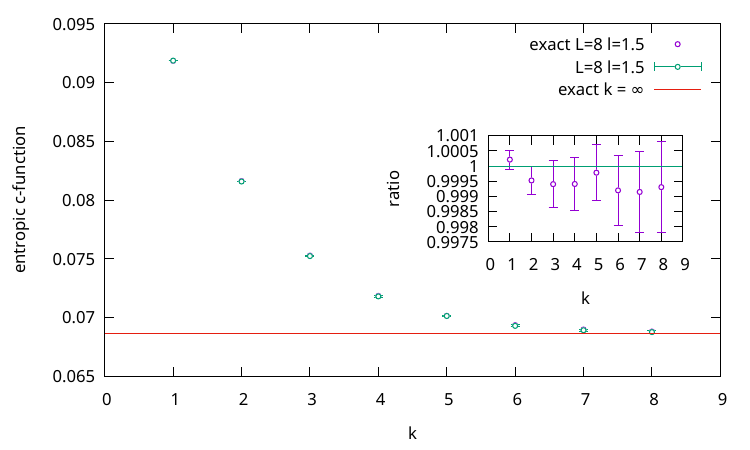}
    \caption{The comparison of the entropic $c$-function obtained in two independent ways: violet data come from the exact transfer matrix method (see Appendix \ref{tran_mat_sect}) and have no statistical uncertainties; green data points were generated with the NIS approach and are shown with their statistical uncertainties. Red horizontal line denotes the exact ground state result obtained by taking $k \to\infty$ in the transfer matrix method. The studied system was the Ising spin chain with $L=8$ spins and the subsystem $A$ had length $l=1.5$. The inset shows the ratio of the values obtained by the two methods.}
    \label{fig:comparison}
\end{figure*}
\end{center}

In what follows we concentrate on the second Rényi entropy, $n=2$, which is the easiest to calculate numerically as the replica trick requires only a duplicated system. In principle, our method applies to any $n>1$, however, in the numerical simulations we are limited by the total number of spins in the system for which the networks can be trained. We first consider the entropic c-functions $C_2$ (\ref{c-funtion_approx}) leaving entropy $S_2$ itself to the end of this section. 

We focus on a single inverse temperature,
$$
\beta = \frac{1}{2}\log\left(1+\sqrt{2}\right) \approx0.4406868,
$$ 
which would correspond to the critical temperature in the classical Ising model at $L\rightarrow \infty$. This allows us to compare our results with analytic predictions from conformal field theory \cite{Calabrese_2009} as well as the numerical results of the same quantity presented in Ref.\cite{Bulgarelli:2023ofi}. The approach works for any value of $\beta$, though. 

To test our method, we start with a small system size of $L=8$, where the numerically exact results can be obtained using the transfer matrix method. For this purpose we express the partition functions $Z$ and $Z_{2}$ as a trace of powers of the transfer matrix. The transfer matrix in this case has size $2^L \times 2^L$ and for small $L$ can be diagonalized exactly. The details of this method can be found in Appendix \ref{tran_mat_sect}. In Fig.~\ref{fig:comparison} we plot the results of entropic c-function Eq.(\ref{c-funtion_approx}) obtained for $L=8$ and $x=l/L=1.5/8$. The horizontal axis represents the time extent of the system in units of  $L$. One can clearly see that when $k$ increases, the entropic c-function converges to the ground state's result (we show this $k=\infty$ result as a red horizontal line, it was obtained using transfer matrix). One can show (for example by investigating the expressions obtained using the transfer matrix approach) that the contributions from excited states decay exponentially with $k$.
We note that the difference between the results obtained using NIS approach (green dots) and the ones obtained using the transfer matrix method (purple dots) cannot be resolved at the scale of the plot. Therefore, we provide an inset where we investigate the ratio of the results obtained using two methods. One notices that NIS gives correct results within errors with per-mile accuracy. The uncertainties of the NIS result are obtained using the bootstrap method. 

\begin{figure*}
\begin{center}
    \includegraphics[width=0.75\textwidth]{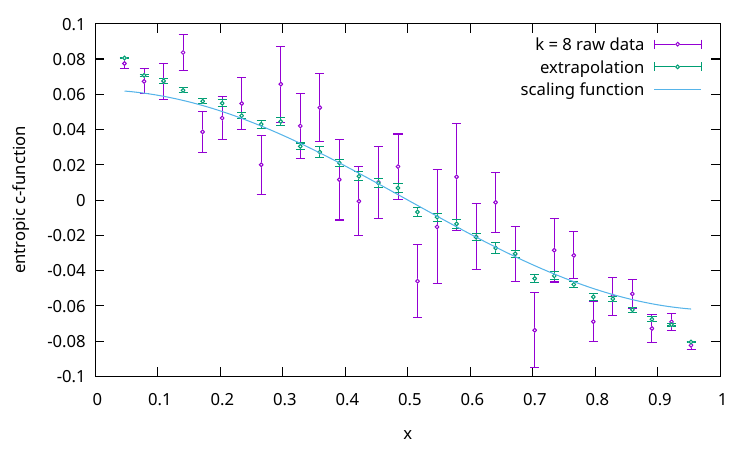}
    \caption{Entropic $c$-function for $L=32$ as a function of $x$. Violet data points correspond to the direct evaluation with $k=8$, whereas the green data points were obtained as a result of the combined extrapolation fit to data with $k=2,\dots,8$ as described in the text. The solid curve shows the theoretical expectation for infinite $L$, the leading term of Eq.~\eqref{eq. scaling function}.}
    \label{fig:improvement}
\end{center}
\end{figure*}

Having tested our method against the transfer matrix approach, we can now move to larger system sizes where the diagonalization of the transfer matrix cannot be efficiently performed. In order to estimate the Rényi entanglement entropy of the ground state, we need to ensure that the path integrals are dominated by that state. From the data in Fig.~\ref{fig:comparison} one can conclude that the time extent of $k=8$ is enough for $L=8$ (see also Ref. \cite{Bulgarelli:2023ofi}, where authors claim that $k=8$ is sufficiently large). For larger systems, we propose to perform an extrapolation with $k\rightarrow \infty$ to enforce the ground state dominance. Before we describe the details of the extrapolation, let us show in Fig.~\ref{fig:improvement} the results of the entropic c-function for $L=32$ as a function of the subsystem size $x=l/L$. The purple dots indicate the results obtained using $k=8$ and we have checked that, within uncertainties, the results for $k=6, 7, 8$ agree with each other. We first observe that the points are distributed such that the anti-symmetry of the entropic c-function, $C_2(1-x)=-C_2(x)$, is satisfied, as it should be for $k \to \infty$. However, the uncertainties are significant, and the expected dependence, called scaling function,  $\frac{1}{16} \cos(\pi x) $, (calculation performed in \cite{Calabrese_2009} for $L \to \infty$) is not visible. In the case of NIS, the size of the uncertainties is directly related to the quality of the training of the neural networks, i.e. how well $q_\theta$ approximates the probability distribution $p$. For large neural networks (with large $L$ and/or $k$), the approximation is worse, resulting in bigger uncertainties. 

In order to reduce the uncertainties, instead of taking just the results for the largest $k$ available, we propose to extrapolate to infinite $k$ using an exponential Ansatz. This allows us to gain statistical precision by using data at several values of $k$. 
Using the property of anti-symmetry of $C_2(x)$ when $k \gg 1$, we can simultaneously fit dependence on $k$ for $x=x_0<0.5$ and $x=1-x_0>0.5$. We use the following model for the fit:
\begin{equation}
	\begin{aligned}
	f(k) & = a + be^{-m k}  \qquad \text{ for $x=x_0$,}   \\
	f(k) &=  -a + \hat be^{-m k} \qquad \text{  for $x=1-x_0$.}
	\end{aligned}
 \label{eq. combined}
\end{equation}
One should note that we also assumed that the data can be described by a single excited state with an energy gap of $m$, 
which is the same for $x=x_0$ and $x=1-x_0$. This is motivated by transfer matrix method expressions. Therefore, we do not expect that the fit will correctly describe data points with $k \sim 1$. With this procedure, we fit four parameters $a, b, \hat b, m$ to 16 points of $C_2(k,x)$ ($k=2,\ldots,8$ and two values of $x$: $x_0$ and $1-x_0$). The  values $C_2(k= \infty, x_0 )$ and $C_2(k= \infty, 1-x_0 )$ are given respectively by $a$ and $-a$. 

\begin{figure*}
\begin{center}
\includegraphics[width=0.75\textwidth]{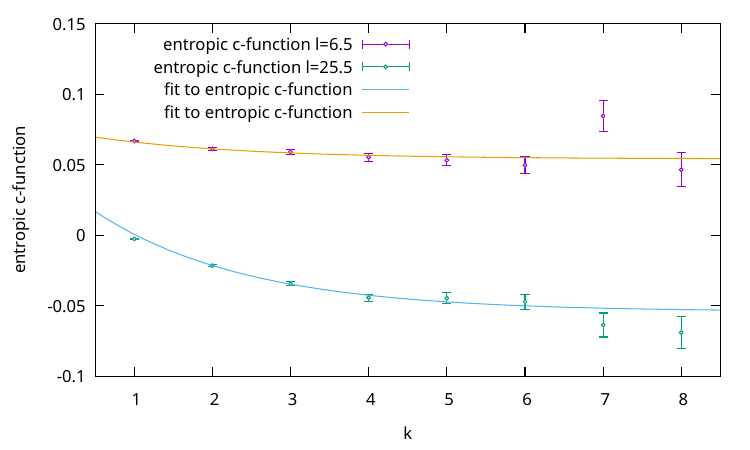}
    \caption{Example of the combined fit of Eq.~\eqref{eq. combined} for $L=32$. The data points at $k=1$ were excluded from the fit because of the large contributions of excited states. }
    \label{fig:result_c}
\end{center}
\end{figure*}

An example of such fit is shown in Fig.~\ref{fig:result_c}, where we show data for two values of $l$: $l=6.5$ and $l=25.5$. Data points are obtained by averaging over all the available statistics. Curves correspond to the combined fit of Eq.\eqref{eq. combined} to all the data points except $k=1$, which exhibits large excited states contributions and is not described by our simple Ansatz. 

The results obtained by the described fitting procedure are shown in Fig.~\ref{fig:improvement} as green dots. One clearly sees that the errors 
were significantly reduced compared to the result using just $k=8$ (purple points).

\begin{figure*}
\begin{center}
    \includegraphics[width=0.75\textwidth]{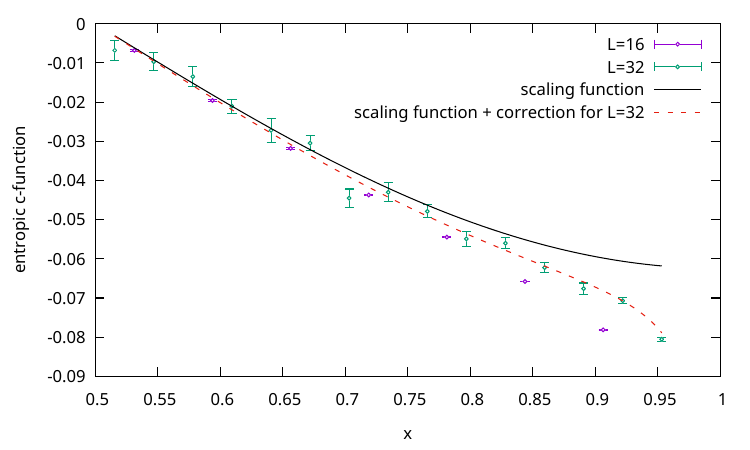}
    \caption{Entropic c-function for $L=16$ and $L=32$ obtained using autoregressive networks (violet and green points). The black solid line (scaling function) denotes the leading term in Eq.~(\ref{eq. scaling function}), $\sim\cos(\pi x)$. Red dashed line denotes full expression (\ref{eq. scaling function}), namely with subleading term $\sim\cot(\pi x) $ included. }
    \label{fig:result_c_comparison}
    \end{center}
\end{figure*}

In Fig.~\ref{fig:result_c_comparison} we show entropic c-function as function of $x$ for $L=16$ and $L=32$. One can see that by increasing $L$ we approach the theoretical result of $\frac{1}{16} \cos(\pi x)$ (named as scaling function) which is obtained in the $L\rightarrow \infty$ limit. Note that in this figure we plot $C_2(x)$ only for $x>0.5$ as the points for $x<0.5$ are not independent due to assumption in the fitting procedure that $C_2(1-x)=-C_2(x)$

To further check our results against the literature, we use the parametrization by Bulgarelli and Panero \cite{Bulgarelli:2023ofi}:
\begin{equation}
\label{eq. scaling function}
    C_2(x) =\frac{1}{16} \cos(\pi x) + \frac{\kappa}{2L}\cot(\pi x)
\end{equation}
where the Authors have fitted $\kappa=0.162$ for $L=32$. The second term is the first-order correction due to the finite size $L$. As is demonstrated in Fig.~\ref{fig:result_c_comparison}, we get a perfect agreement with the results obtained in \cite{Bulgarelli:2023ofi} using the Jarzynski's equality.

\begin{figure*}
\begin{center}
    \includegraphics[width=0.75\textwidth]{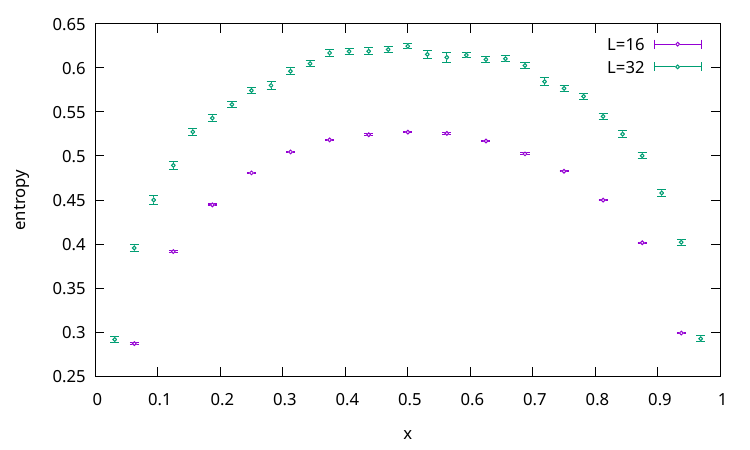}
    \caption{Rényi entropy $S_2$ for $L=16$ and $L=32$ for $k=8$.}
    \label{fig:enter-label}
\end{center}
\end{figure*}

As for the entanglement entropy $S_2$ itself, we plot the result as a function of $x$ in Fig.~\ref{fig:enter-label}. We compare the results of $L=16$ and $L=32$ at $k=8$. The entanglement entropy is maximal at $x=0.5$ (where two subsystems are of the same size) and grows with $L$. We also clearly observe the symmetry property, namely $S_2(x)=S_2(1-x)$

\section{Conclusions} 
\label{sec. summary}

In this contribution, we presented the method of calculating entanglement entropy using generative neural networks. Such neural networks were recently used for simulations of physically interesting quantities in spin systems \cite{2020PhRvE.101b3304N,Bialas:2023fjz,2023CmPhy...6..296B,2024arXiv240216579B} and lattice field theories \cite{PhysRevLett.126.032001,Cranmer:2023xbe}. The key property of those architectures is that they provide explicit probabilities of the generated samples, hence giving access to thermodynamical observables like free energy or entropy. In this manuscript, we have extended the applicability of neural sampling algorithms towards quantum information theory and calculated the entanglement entropy in the one-dimensional quantum Ising model. We compared our results with other methods and found agreement within the statistical uncertainties of our method.

The system sizes that we consider here are rather small and there exist other methods that can be used in this context, resulting in better precision. Nevertheless, our results can be seen as a first step towards developing a new, universal technique for evaluating entanglement entropy. Indeed, the method we used is quite general and can be used in any system where network-based sampling is possible. With the rapid development of deep neural network algorithms, which we are currently witnessing, we believe that our method can be competitive with traditional methods like Quantum Monte Carlo or tensor networks in the future. In particular, the progress in the normalizing flow architectures \cite{Abbott:2023thq,Caselle:2023mvh,PhysRevD.106.074506,Nicoli:2023rcd,Abbott:2024kfc} paves the way toward evaluating entanglement entropy in lattice field theories.

\section*{Acknowledgments}
Computer time allocation 'plgnglft' on the Athena supercomputer hosted by AGH Cyfronet in Krak\'{o}w, Poland was used through the Polish PLGRID consortium. T.S. and D.Z. acknowledge the support of the Polish National Science Center (NCN) Grant No. 2021/43/D/ST2/03375. P.K. acknowledges the support of the Polish National Science Center (NCN) grant No. 2022/46/E/ST2/00346. This research was partially funded by the Priority Research Area Digiworld under the program Excellence Initiative – Research University at the Jagiellonian University in Kraków. We also acknowledge very fruitful discussions with Leszek Hadasz on entanglement entropy.

\appendix

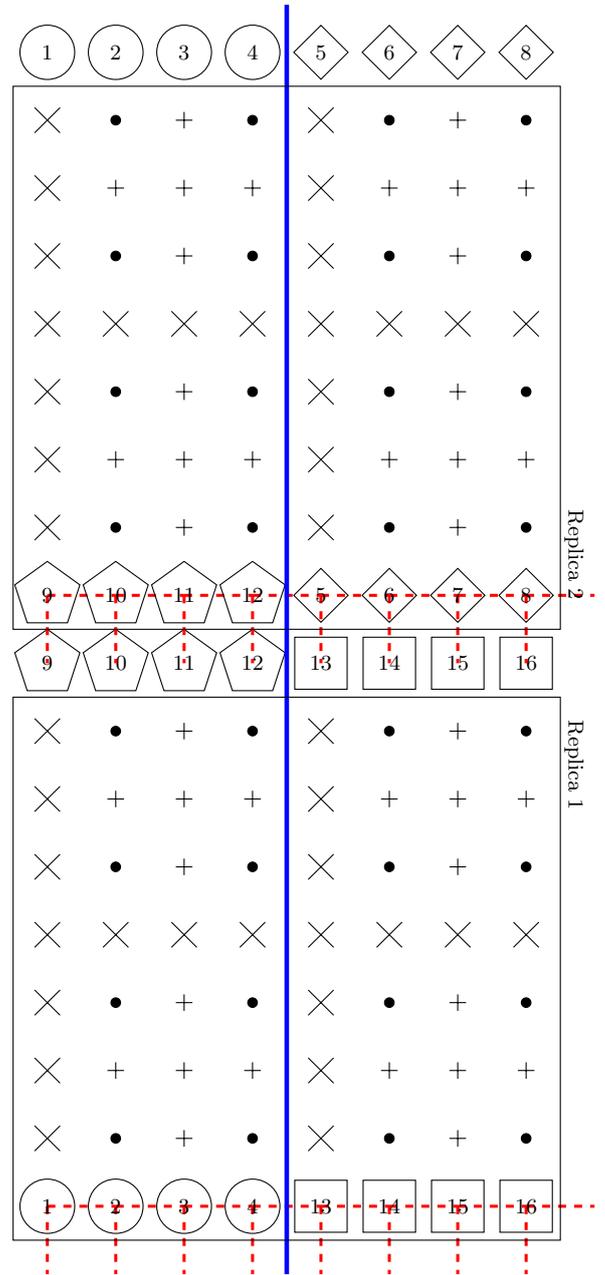
\begin{figure}
\centering
\begin{tikzpicture}[scale=0.9, every node/.style={scale=0.9}]
\newcounter{x}
\newcounter{y}
\newcounter{z}
\setcounter{z}{1}
\forloop{x}{0}{\value{x} < 8}{
    \forloop{y}{0}{\value{y} < 18}{
        \xintifboolexpr{(\value{y} == 0 || \value{y} == 9)}{
            \draw[line width=0.4mm ,red, dashed] (\value{x}, \value{y}) -- (\value{x} + 1, \value{y});
        }
        {}
        \xintifboolexpr{\value{y} == 0 ||\value{y} == 9}{
            \draw[line width=0.4mm, red, dashed] (\value{x}, \value{y}) -- (\value{x}, \value{y} - 1);
            }
        {}

        \xintifboolexpr {(\value{y} == 0 || \value{y} == 17) && \value{x} < 4}
        {
        \setcounter{z}{\value{x} + 1}
        \node at (\value{x}, \value{y}) [circle,draw,minimum size=0.8cm] {\arabic{z}};
        }
        {
        \xintifboolexpr {(\value{y} == 0 || \value{y} == 8) && \value{x} >= 4}
        {
        \setcounter{z}{\value{x} + 9}
        \node at (\value{x}, \value{y}) [regular polygon, regular polygon sides=4,draw] {\arabic{z}};}
        {
        \xintifboolexpr {(\value{y} == 9 || \value{y} == 17) && \value{x} >= 4}
        {
        \setcounter{z}{\value{x} + 1}
        \node at (\value{x}, \value{y}) [diamond, draw] {\arabic{z}};}
        {
        \xintifboolexpr{(\value{y} == 8 || \value{y} == 9) && \value{x} < 4}
        {
        \setcounter{z}{\value{x} + 9}
        \node at (\value{x}, \value{y}) [regular polygon, regular polygon sides=5, minimum size=1.0cm, draw] {\arabic{z}};}
        {
        \xintifboolexpr{(\value{x} == 0 || \value{x} == 4) || \value{y}== 4 || \value{y} == 13}
        {\draw (\value{x}, \value{y}) node[cross=0.2cm, black] {};}
        {
        \xintifboolexpr{(\value{x} == 2 || \value{x} == 6 || \value{y}== 2 || \value{y} == 6 || \value{y} == 11 || \value{y} == 15)}
        {\draw (\value{x}, \value{y}) node[cross=0.1cm, black, rotate=45] {};}
        {\filldraw[black] (\value{x}, \value{y}) circle (2pt);}
        }
        }
        }
        }
        }
    }
}
\node at (7.7, 6.5) [rotate=270] {Replica 1};
\node at (7.7, 9.6) [rotate=270] {Replica 2};
\draw[line width=0.6mm, blue] (3.5, -1.0) -- (3.5, 17.7);
\draw (-0.5, -0.5) rectangle (7.5, 7.5);
\draw (-0.5, 8.5) rectangle (7.5, 16.5);

\end{tikzpicture}
\caption{Sketch of spin division in replica trick. Numbers denote spins fixed first, using the first stage of hierarchy. Repeated number means that the spin is not generated but copied. Spins fixed at other stages of hierarchy are denoted with $\times$, $+$, $\bullet$ marks. Red dashed lines denote connections (interactions) between spins, which were removed when evaluating energy of the configuration. }
\label{replica_trick_details}
\end{figure}

\section{Configurations generation}
\label{han}

In this work, we use hierarchical autoregressive neural networks (HAN) \cite{Bialas:2022qbs} which are modifications of Variational Autoregressive Networks (VAN) \cite{2019PhRvL.122h0602W}. The details of the two architectures can be found for example in Appendix B of \cite{Bialas:2023fjz} and we refer the reader to this appendix for more details. In this contribution we used the same architecture of HAN as described in \cite{Bialas:2023fjz}, with modifications due to the different topology of the configurations, which we shall describe below.

In Fig.~\ref{replica_trick_details} we show a schematic representation of the configuration of two replicas with $k=1$ and $L=8$ "glued together". The blue vertical line separates two subsystems of the spin chain. Different shapes of marks denote the spins fixed at a given stage of the hierarchy:

\begin{enumerate}
	\item  We start with fixing the spins denoted by the numbers from 1 to 16 using an autoregressive network. Then, the values of the spins are copied according to the pattern shown in Fig.~\ref{replica_trick_details}: the repeated numbers denote copies of the spins. This procedure assures the correct topology of the generated configuration: on the left side the replicas are "glued" whereas on the right side each replica is "independent".     
	\item  At the next step of the hierarchy, we generate the spins denoted with "$\times$" mark.
 For this purpose we use autoregressive networks which depend on the spins that were previously fixed, but only those surrounding the given area. These are the networks introduced by the HAN algorithm \cite{Bialas:2022qbs}. 
    \item At this stage, as the $L$ values are powers of $2$, all the remaining spins form squared areas. We proceed with the filling of those areas with the "cross" shapes, which we denote with the "$+$" mark in Fig.~\ref{replica_trick_details}. For the systems with $L>8$, we continue dividing "squared" areas into 4 smaller "squares" using the "cross" shapes, until single spins are left unfixed. 
	\item The hierarchy ends with fixing separated single spins (denoted by "$\bullet$" in Fig.~\ref{replica_trick_details}) - single spins can be drawn using heat bath algorithm, where its probability depends only on the values of the four neighbours (see for example (C1) from Ref.~\cite{Bialas:2023fjz} for the formula). 
\end{enumerate}

The above algorithm, which we described for $k=1$, can be straightforwardly generalized for $k>1$ case. For this purpose, we change the pattern of "$\times$" spins in the second stage of the algorithm - they form a ladder with multiple rungs. 

 One should notice that the number of independent spins generated in total is $k L^2$ for each replica (in Fig.~\ref{replica_trick_details} for  $k=1$, we denote them by frames).

From the technical point of view, the definition of the energy of a configuration has to be adapted to the particular topology imposed by the replica trick and its implementation as shown schematically in Fig.~\ref{replica_trick_details}. In particular, some pairwise interactions between spins, which define the energy of the configuration, need to be adjusted. In Fig.~\ref{replica_trick_details} we denote connections which need to be {\it removed} using red dashed lines. All the other connections (not denoted with red dashed lines), which would appear in the periodic boundary conditions with nearest neighbour interactions, are assumed to be present. The connections in the horizontal direction (for example between spin 1 and 2 in Replica 1) are removed to prevent double counting (the connection between  1 and 2 is already included in the first row). Some of the connections in the vertical direction need to be removed for the same reason (these are connections that join spins with the same number, e.g.~9 and 9). Finally, some vertical connections are removed due to the shape of the replica system (for example, the connection between 5 and 13).

\section{Transfer matrix method}
\label{tran_mat_sect}

For small sizes ($L\leq 10$) we can obtain exact results  using the transfer matrix approach. We define the matrix between two rows of spins $s$ and $s'$ as
\renewcommand{\sp}[1]{s^{\prime {#1}}}
\begin{equation}
\begin{split}
   { \cal A}_{s';s}=\exp\left(
    \beta\sum_{i=0}^{L-1} s^{\prime i}\cdot s^{\prime{i+1}} + \beta\sum_{i=0}^{L-1} s^i\cdot s^{\prime i}
    \right).
\end{split}
\end{equation}
This is a $2^L\times 2^L$ matrix. The partition function of the $L\times M$ spin system is equal to
\begin{equation}
    Z_{L,M}=\operatorname{Tr} { \cal A}^{M}.
\end{equation}

To calculate the partition function $Z_{2}(l)$ of the two replicas, we split the indices of the matrix ${ \cal A}$ into two parts:  $A$ consisting of $l$ spins and $B$ containing the remaining $L-l$ spins. 
$$s=s_A,s_B$$
With such split, the partition function of the two $L\times M$ replicas  connected as depicted in Figure~\ref{replica_trick} can be written as 
\begin{equation}
\begin{split}
    Z_{2} &= \sum_{s_A, s_B, s'_A, s'_B} { \cal A}^{M}_{s'_A,s_B;s'_A,s'_B} { \cal A}^{M}_{s_A,s'_B;s_A,s_B}\\
    &= \sum_{s'_B} { \cal A}^{M}_{B:s_B;s'_B} { \cal A}^{M}_{B:s'_B;s_B}=
    \operatorname{Tr} { \cal A}_B^{2M}
\end{split}    
\end{equation}
where
\begin{equation}
    { \cal A}^{M}_{B:s'_B;s_B}\equiv \sum_{s_A} { \cal A}^{M}_{s_A,s'_B;s_A,s_B}
\end{equation}

Providing  that we can calculate ${ \cal A}^{M}$, such sums can be easily performed numerically for small $L$. The power ${ \cal A}^M$ can be calculated by noting that 
\begin{equation}\label{eq:A-T}
{ \cal A}_{s';s} = p_{s'}{ \cal T}_{s';s}p^{-1}_{s},
\end{equation}
where 
\begin{equation}
\begin{split}
    { \cal T}_{s';s}=\exp\Biggl(
    \beta\frac{1}{2}&\sum_{i=0}^{L-1} \left(s^i\cdot s^{i+1}+\sp{i}\cdot \sp{i+1}\right)+\\
    &\beta\sum_{i=0}^{L-1} s_i\cdot s'_i
    \Biggr)
\end{split}
\end{equation}
and 
$$p_{s'}=\exp\left(
    \beta\frac{1}{2}\sum_{i=0}^{L-1} \sp{i}\cdot \sp{i+1} \right).
$$    
Please note that there is no implied summation in \eqref{eq:A-T}. Formula \eqref{eq:A-T} also entails 
\begin{equation}\label{eq:A-power}
{ \cal A}^M_{s';s} = p_{s'}{ \cal T}^M_{s';s}p^{-1}_{s}    
\end{equation}
Contrary to ${ \cal A}$ matrix ${ \cal T}$ is symmetric and can be written in the form
$${ \cal T}^{M} = { \cal P}\cdot { \cal D}^{M} \cdot { \cal P}^T$$
where ${ \cal D}$ is the diagonal eigenvalues matrix, and ${ \cal P}$ is the  orthogonal  matrix of eigenvectors. From \eqref{eq:A-power} we finally  obtain
$${ \cal A}^{k L}_{s',s} = p_{s'} ({ \cal P}\cdot { \cal D}^{ k L} \cdot { \cal P}^T)_{s',s} p^{-1}_{s}.$$
All these calculations can be easily performed using \texttt{NumPy} library. The implementation can be found in the accompanying notebook.

\bibliographystyle{ieeetr}
\bibliography{references2}

\begin{thebibliography}{10}

\bibitem{Horodecki2009}
R.~Horodecki, P.~Horodecki, M.~Horodecki, and K.~Horodecki, ``Quantum entanglement,'' {\em Reviews of Modern Physics}, vol.~81, pp.~865--942, 6 2009.

\bibitem{Calabrese_2009}
P.~Calabrese and J.~Cardy, ``Entanglement entropy and conformal field theory,'' {\em Journal of Physics A: Mathematical and Theoretical}, vol.~42, p.~504005, dec 2009.

\bibitem{PhysRevLett.104.157201}
M.~B. Hastings, I.~Gonz\'alez, A.~B. Kallin, and R.~G. Melko, ``Measuring renyi entanglement entropy in quantum monte carlo simulations,'' {\em Phys. Rev. Lett.}, vol.~104, p.~157201, Apr 2010.

\bibitem{Humeniuk:2012xg}
S.~Humeniuk and T.~Roscilde, ``{Quantum Monte Carlo calculation of entanglement Renyi entropies for generic quantum systems},'' {\em Phys. Rev. B}, vol.~86, p.~235116, 2012.

\bibitem{PhysRevB.90.125105}
D.~J. Luitz, X.~Plat, N.~Laflorencie, and F.~Alet, ``Improving entanglement and thermodynamic r\'enyi entropy measurements in quantum monte carlo,'' {\em Phys. Rev. B}, vol.~90, p.~125105, Sep 2014.

\bibitem{Zhao_2022}
J.~Zhao, B.-B. Chen, Y.-C. Wang, Z.~Yan, M.~Cheng, and Z.~Y. Meng, ``Measuring rényi entanglement entropy with high efficiency and precision in quantum monte carlo simulations,'' {\em npj Quantum Materials}, vol.~7, June 2022.

\bibitem{PhysRevLett.124.110602}
J.~D'Emidio, ``Entanglement entropy from nonequilibrium work,'' {\em Phys. Rev. Lett.}, vol.~124, p.~110602, Mar 2020.

\bibitem{Bulgarelli:2023ofi}
A.~Bulgarelli and M.~Panero, ``{Entanglement entropy from non-equilibrium Monte Carlo simulations},'' {\em JHEP}, vol.~06, p.~030, 2023.

\bibitem{Eisert_2013}
J.~Eisert, ``Entanglement and tensor network states,'' {\em Modeling and Simulation}, vol.~3, p.~520, 2013.

\bibitem{Okunishi_2022}
K.~Okunishi, T.~Nishino, and H.~Ueda, ``Developments in the tensor network {\textemdash} from statistical mechanics to quantum entanglement,'' {\em Journal of the Physical Society of Japan}, vol.~91, jun 2022.

\bibitem{Kogut1979}
J.~B. Kogut, ``An introduction to lattice gauge theory and spin systems,'' {\em Reviews of Modern Physics}, vol.~51, p.~659, 10 1979.

\bibitem{2019PhRvL.122h0602W}
D.~{Wu}, L.~{Wang}, and P.~{Zhang}, ``{Solving Statistical Mechanics Using Variational Autoregressive Networks},'' {\em Phys. Rev. Lett.}, vol.~122, p.~080602, Mar. 2019.

\bibitem{Bialas:2021bei}
P.~Bia\l{}as, P.~Korcyl, and T.~Stebel, ``{Analysis of autocorrelation times in neural Markov chain Monte Carlo simulations},'' {\em Phys. Rev. E}, vol.~107, no.~1, p.~015303, 2023.

\bibitem{Bialas:2022qbs}
P.~Bia\l{}as, P.~Korcyl, and T.~Stebel, ``{Hierarchical autoregressive neural networks for statistical systems},'' {\em Comput. Phys. Commun.}, vol.~281, p.~108502, 2022.

\bibitem{Bialas:2022bdl}
P.~Bia\l{}as, P.~Czarnota, P.~Korcyl, and T.~Stebel, ``{Simulating first-order phase transition with hierarchical autoregressive networks},'' {\em Phys. Rev. E}, vol.~107, no.~5, p.~054127, 2023.

\bibitem{2020PhRvE.101b3304N}
K.~A. {Nicoli}, S.~{Nakajima}, N.~{Strodthoff}, W.~{Samek}, K.-R. {M{\"u}ller}, and P.~{Kessel}, ``{Asymptotically unbiased estimation of physical observables with neural samplers},'' {\em Phys. Rev. E}, vol.~101, p.~023304, Feb. 2020.

\bibitem{Bialas:2023fjz}
P.~Bia\l{}as, P.~Korcyl, and T.~Stebel, ``{Mutual information of spin systems from autoregressive neural networks},'' {\em Phys. Rev. E}, vol.~108, no.~4, p.~044140, 2023.

\bibitem{2023CmPhy...6..296B}
I.~{Biazzo}, ``{The autoregressive neural network architecture of the Boltzmann distribution of pairwise interacting spins systems},'' {\em Communications Physics}, vol.~6, p.~296, Dec. 2023.

\bibitem{2024arXiv240216579B}
I.~{Biazzo}, D.~{Wu}, and G.~{Carleo}, ``{Sparse Autoregressive Neural Networks for Classical Spin Systems},'' {\em arXiv e-prints}, p.~arXiv:2402.16579, Feb. 2024.

\bibitem{PhysRevLett.126.032001}
K.~A. Nicoli, C.~J. Anders, L.~Funcke, T.~Hartung, K.~Jansen, P.~Kessel, S.~Nakajima, and P.~Stornati, ``Estimation of thermodynamic observables in lattice field theories with deep generative models,'' {\em Phys. Rev. Lett.}, vol.~126, p.~032001, Jan 2021.

\bibitem{Cranmer:2023xbe}
K.~Cranmer, G.~Kanwar, S.~Racani\`ere, D.~J. Rezende, and P.~E. Shanahan, ``{Advances in machine-learning-based sampling motivated by lattice quantum chromodynamics},'' {\em Nature Rev. Phys.}, vol.~5, no.~9, pp.~526--535, 2023.

\bibitem{Abbott:2023thq}
R.~Abbott {\em et~al.}, ``{Normalizing flows for lattice gauge theory in arbitrary space-time dimension},'' 5 2023.

\bibitem{Caselle:2023mvh}
M.~Caselle, E.~Cellini, and A.~Nada, ``{Sampling the lattice Nambu-Goto string using Continuous Normalizing Flows},'' {\em JHEP}, vol.~02, p.~048, 2024.

\bibitem{PhysRevD.106.074506}
R.~Abbott, M.~S. Albergo, D.~Boyda, K.~Cranmer, D.~C. Hackett, G.~Kanwar, S.~Racani\`ere, D.~J. Rezende, F.~Romero-L\'opez, P.~E. Shanahan, B.~Tian, and J.~M. Urban, ``Gauge-equivariant flow models for sampling in lattice field theories with pseudofermions,'' {\em Phys. Rev. D}, vol.~106, p.~074506, Oct 2022.

\bibitem{Nicoli:2023rcd}
K.~A. Nicoli, C.~J. Anders, L.~Funcke, K.~Jansen, S.~Nakajima, and P.~Kessel, ``{NeuLat: a toolbox for neural samplers in lattice field theories},'' {\em PoS}, vol.~LATTICE2023, p.~286, 2024.

\bibitem{Abbott:2024kfc}
R.~Abbott, A.~Botev, D.~Boyda, D.~C. Hackett, G.~Kanwar, S.~Racani\`ere, D.~J. Rezende, F.~Romero-L\'opez, P.~E. Shanahan, and J.~M. Urban, ``{Applications of flow models to the generation of correlated lattice QCD ensembles},'' {\em Phys. Rev. D}, vol.~109, no.~9, p.~094514, 2024.

\end{thebibliography}

\end{document}